\documentclass[preprint,superscriptaddress,showpacs,tightenlines]{revtex4}
\usepackage{amssymb}
\usepackage{amsmath}
\usepackage{graphicx,bm}

\input{tcilatex}

\begin{document}

\title{Conductance of a tunnel point-contact of noble metals in the presence of
a single defect }
\author{Ye.S. Avotina}
\affiliation{B.I. Verkin Institute for Low Temperature Physics and Engineering, National
Academy of Sciences of Ukraine, 47, Lenin Ave., 61103, Kharkov,Ukraine.}
\affiliation{Kamerlingh Onnes Laboratorium, Universiteit Leiden, Postbus 9504, 2300
Leiden, The Netherlands.}
\author{Yu.A. Kolesnichenko}
\affiliation{B.I. Verkin Institute for Low Temperature Physics and Engineering, National
Academy of Sciences of Ukraine, 47, Lenin Ave., 61103, Kharkov,Ukraine.}
\affiliation{Kamerlingh Onnes Laboratorium, Universiteit Leiden, Postbus 9504, 2300
Leiden, The Netherlands.}
\author{S.B. Roobol}
\affiliation{Kamerlingh Onnes Laboratorium, Universiteit Leiden, Postbus 9504, 2300
Leiden, The Netherlands.}
\author{J.M. van Ruitenbeek}
\affiliation{Kamerlingh Onnes Laboratorium, Universiteit Leiden, Postbus 9504, 2300
Leiden, The Netherlands.}

\begin{abstract}
In paper [1] (Avotina \textit{et al.}\ Phys. Rev. B \textbf{74},
085411 (2006)) the effect of Fermi surface anisotropy to the
conductance of a tunnel point contact, in the vicinity of which a
single point-like defect is situated, has been investigated
theoretically. The oscillatory dependence of the conductance on
the distance between the contact and the defect has been found for
a general Fermi surface geometry. In this paper we apply the
method developed in [1] to the calculation of the conductance of
noble metal contacts. An original algorithm, which enables the
computation of the conductance for any parametrically given Fermi
surface, is proposed. On this basis a pattern of the conductance
oscillations, which can be observed by the method of scanning
tunneling microscopy, is obtained for different orientations of
the surface for the noble metals.
\end{abstract}

\maketitle

\bigskip The scanning tunnelling microscope (STM) method enables to observe
and investigate quantum interference phenomena concerned with
electron scattering by single defects. One of them is Friedel-like
oscillations of the differential tunneling conductance $G$
measured by STM around the defect. It is known that electrons of
the surface states on the (111) surfaces of the noble metals Au,
Ag, and Cu form a quasi-two-dimensional electron gas which is
confined at the crystal surface. These electrons are scattered by
surface defects, e.g. impurity atoms, adatoms, or step edges, and
the STM conductance exhibits oscillatory patterns originating from
an interference between the principal wave that is directly
transmitted through the contact
and the partial wave that is scattered by the contact and the defect \cite%
{Cromme,Koles,Knorr,Burge}. The period of the conductance
oscillations depends on a distance from the contact to the defect
$r_{0}$ and double the Fermi wave vector $2k_{F}$. A similar
dependence can result from the scattering of bulk electron states
by subsurface defects \cite{babble,Quaas,Wend}. It was found that
the oscillatory pattern obtained by STM reflects the anisotropy of
the Fermi surface (FS), i.e. the value of the vector
$\mathbf{k}_{F}$ depends on the direction in a plane of the sample
surface, and surface Fermi contours can be
determined by Fourier transform of the STM image \cite{Ber,Fourier,Petersen}%
. Particularly, in Ref. \cite{Petersen} the contour related to the 'neck' of
the bulk FS that for Cu (111) and Au (111) surfaces had been observed.

In the papers \cite{Avotina1,Avotina2,Avotina3} the effect of
quantum interference of electron waves which are scattered by
single defects below a metal surface to the conductance of a
tunnel point-contact has been investigated theoretically. It has
been shown \cite{Avotina1} that the dependence of $G$ on an
applied voltage measure $V$ can be used for the
determination of defect positions below a metal surface. In Ref. \cite%
{Avotina2} we have analyzed the conductance of a tunnel
point-contact in the presence of a defect located inside the bulk
for metals with arbitrary FS. In the quasiclassical approximation
the conductance of the contact had been found. The general formula
was illustrated for two non-spherical shapes for the FS: the
ellipsoid and the corrugated cylinder (open surface). These
relatively simple models of FS make it possible to get analytical
expressions for the conductance and to analyze the main
manifestations of the FS anisotropy: 'necks', inflection lines
etc. In order time to compare the theoretical results with
experiment it is necessary to calculate the conductance for the
real model of the FS of a specific metal. In this paper we present
such calculations for noble metals.

We consider as a model for our system a nontransparent interface separating
two metal half-spaces, in which there is an orifice (contact) of radius $%
a\ll \lambda _{\mathrm{F}}$ ($\lambda _{\mathrm{F}}$ is a
characteristic Fermi wave length). The potential barrier in the
plane of the contact is taken to be a delta function with a large
amplitude $U$ (the transmission coefficient of electron tunneling
through the barrier is small, $T\approx \left( \hbar
v_{\mathrm{F}}/U\right) ^{2}\ll 1,$ $v_{\mathrm{F}}$ is a Fermi
velocity). At the distance $r_{0}\gg \lambda _{\mathrm{F}}$ from
the contact a point-like defect, which is described by a short
range potential, is placed. The interaction of electrons with the
defect is taken into account in the framework of perturbation
theory with the constant of this interaction $g$.
We also assume the applied bias $eV$ is much smaller than Fermi energy, $%
\varepsilon _{\mathrm{F}}$. The conductance of the contact is calculated in
linear approximation in the transmission coefficient $T,$ the constant $%
g $ and the voltage $V$ by the method developed in Refs. \cite{Avotina1,KMO}%
. Under the listed assumptions a general formula for the
conductance had been derived in Ref. \cite{Avotina2}:
\begin{equation}
G=G_{0}\left( 1-\widetilde{g}\sum_{s,s^{\prime }}\func{Re}\Lambda _{s}^{%
\mathrm{as}}\left( \mathbf{r}_{0}\mathbf{,}\varepsilon _{\mathrm{F}}\right)
\func{Im}\Lambda _{s^{\prime }}^{\mathrm{as}}\left( \mathbf{r}_{0}\mathbf{,}%
\varepsilon _{\mathrm{F}}\right) \right) .  \label{G}
\end{equation}%
Here $G_{0}\sim T$ is the conductance of the tunnel point-contact
without defect, $\widetilde{g}\approx gm^{\ast
2}v_{\mathrm{F}}/\hbar ^{3}$ is a dimensionless constant of
electron-impurity interaction ($m^{\ast }$ is the effective
electron mass). The function $\Lambda _{s}^{\mathrm{as}}\left(
\mathbf{r,}\varepsilon \right) $ defines the asymptote of the wave
function $\psi \left( \mathbf{r}\right) \sim \sqrt{T}\Lambda
^{\mathrm{as}}\left( \mathbf{r}\right) $ of the electrons
transmitted through the contact at large distances $r\gg \lambda
_{\mathrm{F}}$ from the contact. For points in the
momentum space, for which the Gaussian curvature $K\left( \varepsilon ,%
\mathbf{p}_{t}\right) $ of the FS $\varepsilon \left( \mathbf{p}%
_{t},p_{z}\right) =\varepsilon _{\mathrm{F}}$ ($z$ is directed
along the contact axis, $\mathbf{p}_{t}$ and $p_{z}$ are
components of the momentum tangential and perpendicular to the
interface) is not equal to zero, the
function $\Lambda _{s}^{\mathrm{as}}\left( \mathbf{r,}\varepsilon _{\mathrm{F%
}}\right) $ is given by
\begin{equation}
\Lambda _{s}^{\mathrm{as}}\left( \mathbf{r,}\varepsilon _{\mathrm{F}}\right)
=\frac{\cos \vartheta }{2\pi \hbar r\sqrt{\left\vert K\right\vert }}\exp %
\left[ i\Gamma +i\frac{\pi }{4}\text{sgn}\left( \frac{\partial
^{2}p_{z}^{\left( +\right) }}{\partial p_{x}^{2}}\right) \left( 1+\text{sgn}%
K\right) \right] _{\mathbf{p}_{t}=\mathbf{p}_{t,s}^{\left( \mathrm{st}%
\right) }},  \label{lam}
\end{equation}%
where $\Gamma \left( \mathbf{p}_{t},\mathbf{r}\right) $ is the phase
accumulated over the path travelled by the electron between the contact and
the point $\mathbf{r}$,
\begin{equation}
\Gamma \left( \mathbf{p}_{t},\mathbf{r}\right) =\frac{1}{\hbar }\left(
\mathbf{p}_{t}\mathbf{\rho }+p_{z}^{\left( +\right) }\left( \mathbf{p}%
_{t}\right) z\right) ,  \label{phase}
\end{equation}%
$p_{z}^{\left( +\right) }\left( \varepsilon _{\mathrm{F}},\mathbf{p}%
_{t}\right) $ is the root of the equation $\varepsilon \left( \mathbf{p}%
_{t},p_{z}^{\left( +\right) }\right) =\varepsilon _{\mathrm{F}}$
corresponding to a wave with a $z$-component of the velocity $v_{z}^{\left(
+\right) }\left( \varepsilon _{\mathrm{F}},\mathbf{p}_{t}\right) >0$, and $%
\cos \vartheta (\mathbf{r})=z/r$ is the angle between the vector $\mathbf{r}$
and the $z$ axis. The momenta $\mathbf{p}_{t}=\mathbf{p}_{t,s}^{\left(
\mathrm{st}\right) }$ ($s=1,2...$) are defined by the equation,
\begin{equation}
\left. \frac{\partial \Gamma }{\partial \mathbf{p}_{t}}\right\vert _{\mathbf{%
p}_{t}=\mathbf{p}_{t,s}^{\left( \mathrm{st}\right) }}=0.  \label{st_point}
\end{equation}%
Originally, $\mathbf{p}_{t}=\mathbf{p}_{t,s}^{\left(
\mathrm{st}\right) }$ are the stationary phase points of the
integral wave function \cite{Avotina2}.
These projections of the momentum correspond to the velocities $\mathbf{v}%
\left( \varepsilon ,\mathbf{p}_{t,s}^{\left( \mathrm{st}\right) }\right)
\parallel \mathbf{r,}$ i.e. at large distances from the contact the electron
wave function for a certain direction $\mathbf{r}$ is defined by those
points on the FS for which the electron group velocity is parallel to $%
\mathbf{r}$ \cite{Avotina2,Kosevich}. If the curvature of the FS
changes sign, Eq.~(\ref{st_point}) has more than one solution
($s=1,2...$). It may also occur that Eq.~(\ref{st_point}) does not
have any solution for given directions of the vector $\mathbf{r}$,
and the electrons cannot propagate along these directions
\cite{LAK}.

At the stationary phase points the curvature $K\left( \varepsilon ,\mathbf{p}%
\right) $ can be written as
\begin{equation}
K_{0}\left( \varepsilon ,\mathbf{n}\right) =\left[ \frac{1}{\left\vert
\mathbf{v}\right\vert ^{2}}\sum_{i,k=x,y,z}A_{ik}n_{i}n_{k}\right] _{\mathbf{%
\ p}_{t}=\mathbf{p}_{t}^{\left( \mathrm{st}\right) }},  \label{Gaus0}
\end{equation}%
where $A_{ik}=\frac{\partial \det \left( \mathbf{m}^{-1}\right) }{\partial
m_{ik}^{-1}\left( \mathbf{p}\right) }$ is the algebraic adjunct of the
element
\begin{equation}
m_{ik}^{-1}\left( \mathbf{p}\right) =\frac{\partial ^{2}\varepsilon }{%
\partial p_{i}\partial p_{k}}  \label{m^(-1)}
\end{equation}%
of the inverse mass matrix $\mathbf{m}^{-1}$ \cite{Korn}; $n_{i}$ are
components of the unit vector $\mathbf{n}=\mathbf{r}/r.$

For those points at which $K_{0}=0$ the amplitude of the electron
wave function in a direction of zero Gaussian curvature is larger
than for other directions. This results in an enhanced current
flow near the cone surface defined by the condition $K_{0}=0$
\cite{Kosevich}. If the FS is open, there are directions along
which the electrons can not move at all. These properties of the
wave function manifest itself in an oscillatory part of the
conductance (\ref{G}): 1) The amplitude of oscillations is maximal
if the direction from the contact to the defect corresponds to the
electron velocity belonging to an inflection line. 2) There are no
oscillations of $G$ if this direction belong to cones, in which
the electron motion is forbidden.

Further calculations requires the information about the FS,
$\varepsilon \left( \mathbf{p}\right) =\varepsilon _{\mathrm{F}}.$
We use the parameterization of the FS of the noble metals copper,
silver and gold from \cite{FS}
\begin{eqnarray}
\varepsilon (\mathbf{p}) &=&\alpha \left[ -3+\text{cos}\frac{p_{x}a}{2\hbar }%
\text{cos}\frac{p_{y}a}{2\hbar }+\text{cos}\frac{p_{y}a}{2\hbar }\text{cos}%
\frac{p_{z}a}{2\hbar }+\text{cos}\frac{p_{z}a}{2\hbar }\text{cos}\frac{p_{x}a%
}{2\hbar }+\right.  \label{E(k)} \\
&&\left. r\left( -3+\text{cos}\frac{p_{x}a}{\hbar }+\text{cos}\frac{p_{y}a}{%
\hbar }+\text{cos}\frac{p_{z}a}{\hbar }\right) \right] .  \notag
\end{eqnarray}

This parameterization is accurate up to 99\%. The value of the constants are $%
r=0.0995$, and $\varepsilon /\alpha $ $=3.63$, and $a$ is
different for each metal. For copper, silver and gold $a=0.361$,
$a=0.408$ and $a=0.407nm$, respectively. The Fermi energy of
copper is 7.00 eV, for silver 5.49 eV and for gold it is 5.53 eV.

The FS (\ref{E(k)}) has the BCC\ symmetry. It basically looks like a sphere
with 8 'necks' positioned at the 8 vertices of a cube (Fig. 1). The central
part of the surface (`belly') has a positive curvature $K>0$ while the ends
near the Brillouin zone boundary (`necks') have negative curvature. The size
of the 'necks' and the curvature of the spherical areas are slightly
different for each noble metal. In the regions of 'necks' there are the
inflection lines, at which the curvature $K=0.$

In Eq. (\ref{E(k)}) for the FS, the $x$, $y$ and $z$ directions
correspond to a [100] direction, and the $xy$ interface plane is
therefore a (100) crystal plane. To align the $xy$ plane with a
(110) plane, the FS is rotated by $\pi /4$ along the $x$ or $y$
axis. For the (111) orientation, the total rotation consists of a
rotation of $\pi /4$ along the $z$-axis, followed by a rotation of
arcsin$(1/\sqrt{3})$ along the $x$ or $y$ axis.

A direct way to find solutions $\mathbf{p}_{t}=\mathbf{p}_{t,s}^{\left(
\mathrm{st}\right) }$ of Eq. (\ref{st_point}) numerically for a certain
position of the defect $\mathbf{r}_{0}$ and calculate $\Lambda _{s}^{\mathrm{%
as}}\left( \mathbf{r}_{0}\mathbf{,}\varepsilon _{\mathrm{F}}\right) $ (\ref%
{lam}) is not most suitable. Instead, the result that $\mathbf{p}_{t}=\mathbf{p}%
_{t,s}^{\left( \mathrm{st}\right) }$ corresponds to the direction of the
electron velocity along the direction from the contact to the defect can be
used. We start with a point $\mathbf{p}$ on the FS, calculate the value of $%
\Lambda _{s}^{\mathrm{as}}\left( \mathbf{r,}\varepsilon _{\mathrm{F}}\right)
$ for every point $\mathbf{r\parallel v}=\partial \varepsilon \left( \mathbf{p%
}\right) /\partial \mathbf{p}$ in the real space, and then repeat
this for all points on the FS. Next, it is easy to perform the
summation over all points on the FS, in which $\mathbf{r\parallel
v}$ to obtain the conductance. This idea is shown schematically in
Fig. 1.

\begin{figure}[tbp]
\includegraphics[width=10cm,angle=0]{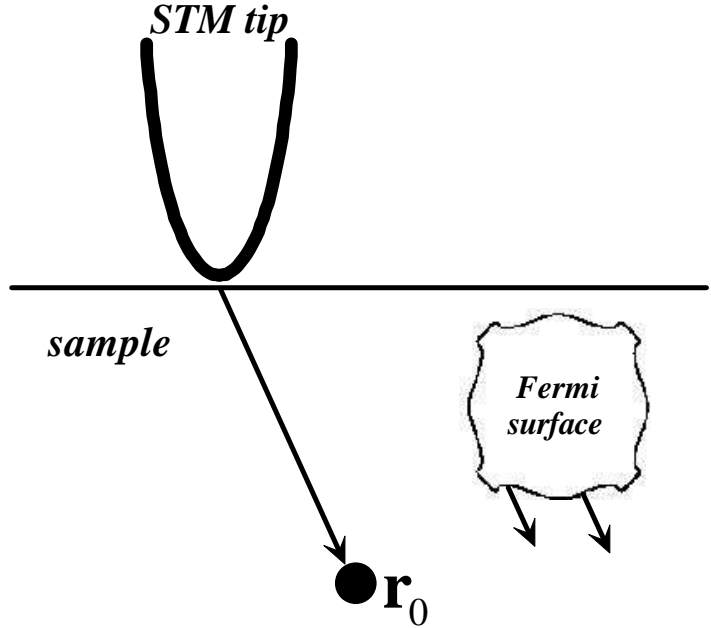}
\caption{The main contributions to the conductance oscillations
caused by a defect at $\mathbf{r}_{o}$ come from the points on the
Fermi surface where the normal vector (velocity vector) points in
the same direction as the vector $\mathbf{r}_{o}.$} \label{fig1}
\end{figure}

Strictly speaking, the asymptotic Eq. (\ref{G}) is correct at
$a\ll \lambda _{\mathrm{F}},$ $r_{0}\gg \lambda _{\mathrm{F}}.$
However, as it was shown in Ref. \cite{Avotina2} from a comparison
of the exact result for the ellipsoidal FS with asymptotic
expression (\ref{G}), Eq. (\ref{G}) describes the conductance
qualitatively correctly for $a<\lambda _{\mathrm{F}}$ and
distances $r_{0}$ of a few $\lambda _{\mathrm{F}}.$ The other
point is that at the inflection lines, which defines the
classically unacceptable regions, the curvature $K=0.$ As it was
shown in Ref. \cite{Avotina2}, for such directions of the vector
$\mathbf{r}_{0}$ the amplitude of the conductance oscillations
increases remains finite. Below we restrict ourselves to the
condition $K\neq 0$ and do not approach the inflection lines to a
distance for which the second term in the Eq. (\ref{G}) becomes of
the order of unity.

\begin{figure}[tbp]
\includegraphics[width=10cm,angle=0]{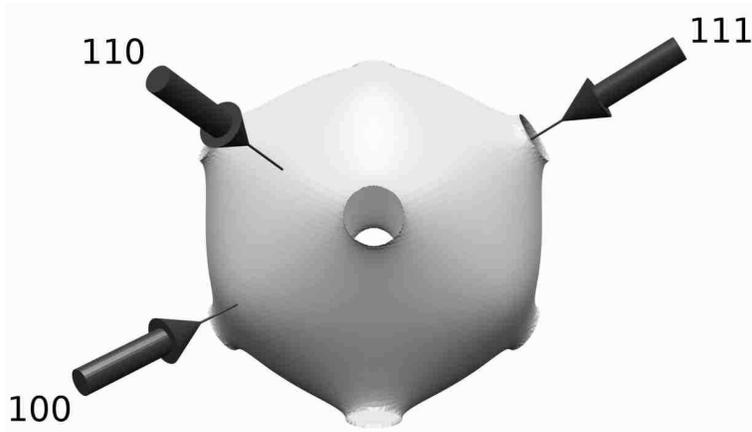}
\caption{The orientation of the Fermi surface relative to the
contact axis for three principal lattice orientations.}
\label{fig2}
\end{figure}

We present the result of computations for three different
crystallographic orientations (Fig. 2). The conductance as a
function of the contact position for a defect in a noble metal at
various depths are plotted in Figs. 3-5 for the (100), (110) and
(111) lattice orientations respectively. For each of the lattice
orientations, the graphs have the symmetries of that particular
orientation of the FS. In all figures 'dead' regions, in which
there are no conductance oscillations, can be seen. These regions
originate from the 'necks' of the FS and their edges are defined
by the inflection lines. In our plots the edges are abrupt. In
reality there is a smooth change from a maximum to a zero of
amplitude of the oscillations in the 'dead' regions. This change
can not be described by Eq.(\ref{G}), and a numerical solution of the Schr\"{o}%
dinger equation with energy-momentum relation (\ref{E(k)}) must be
used. However, the problem becomes much more complicated, while it
does not give any additional physical information. The rings of
high amplitude conductance oscillations
have already been reported in experiments on Ag and Cu (111) surfaces \cite%
{Quaas}.

\begin{figure}[tbp]
\includegraphics[width=12cm,angle=0]{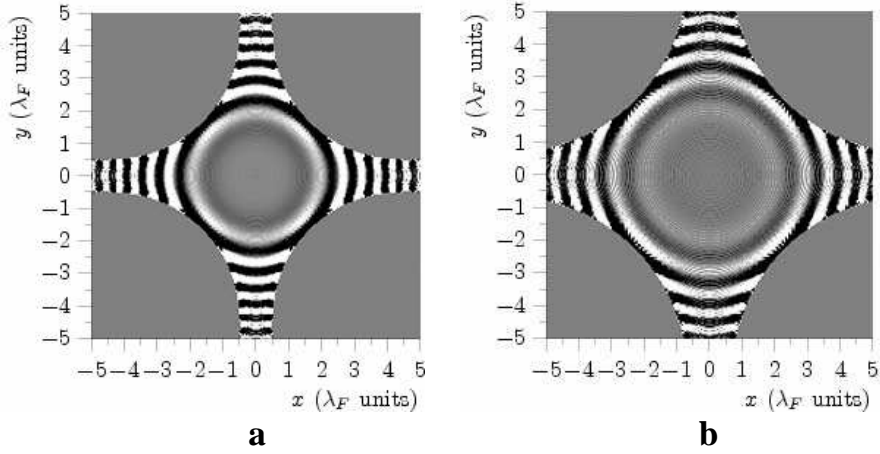}
\caption{Conductance $G$ as a function of the contact position for
a defect at the origin at depths 5$\lambda_{F}$ (a) and
7$\lambda_{F}$ (b) for a (100) interface plane. The $x$ and $y$
directions both correspond to $\left\langle 100\right\rangle $
directions. The conductance is plotted in gray scale, where the
color of the 'dead' regions corresponds to the conductance value
in absence the defect $G=G_{0}$, positive addition to $G$ is white
and negative is black.} \label{fig3}
\end{figure}

\begin{figure}[tbp]
\includegraphics[width=12cm,angle=0]{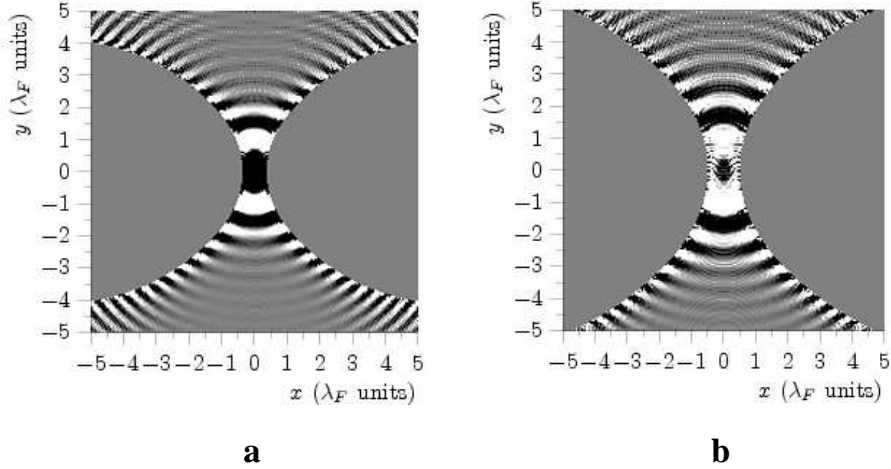}
\caption{Same as Fig. 3, but for a (110)
interface plane. The $x$ and $y$ directions correspond to $[001]$ and $[1%
\overline{1}0]$ directions respectively.} \label{fig4}
\end{figure}

\begin{figure}[tbp]
\includegraphics[width=12cm,angle=0]{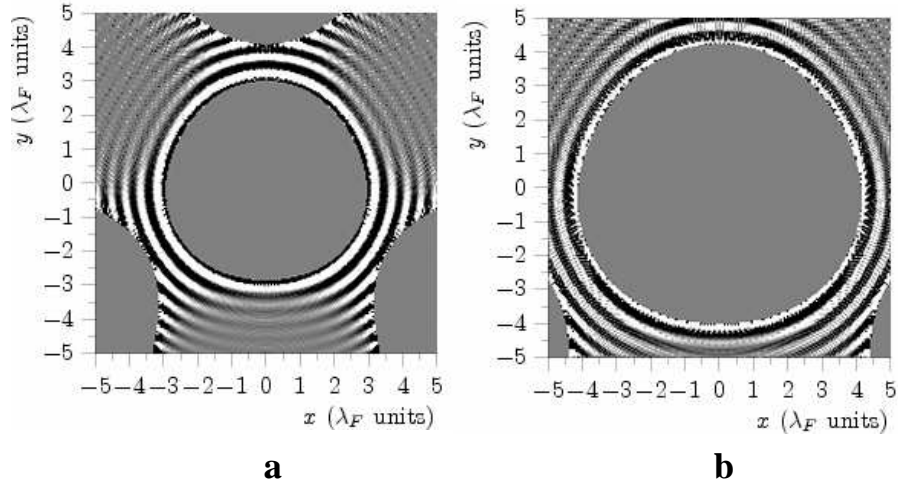}
\caption{Same as Fig. 3, but for a (111)
interface plane. The $x$ and $y$ directions correspond to $[11\overline{2}]$%
and $[1\overline{1}0]$ directions respectively.} \label{fig5}
\end{figure}

From Figs. 3-5 it can be seen that the interference pattern of the
conductance oscillations, in particular the size and appearance of
'dead' regions depend on the depth of the defect. These
characteristics of the part of the conductance related to the
scattering by the defect contains the information about the
position of the defect. For all orientations of the metal surface
the defect position in the plane of the surface corresponds to a
center of symmetry. The depth can be found in the following way:
The orientation of the 'neck' axes defines the axes of the cones,
in which there are no scattered electrons. Vertexes of the cones
coincide with the defect. If the contact is situated in a point,
which belong to a sectional plane of the cones by a surface plane,
the conductance of the contact is equal to its value without the
defect (we called these 'dead' regions). A rough estimation of the
defect depth can be obtain, if we use the approximation of a cone
of revolution with an opening angle $2\gamma .$ For example, in
Fig.5 the radius $R$ of the central 'dead' region is defined by
the equality $R=z_{0}\cot \left( \gamma \right)$,($\gamma \approx
30^{0}$) \cite{Quaas}. Using a fiting of experimental results with
theoretical calculations in the framework our method enables one
to find the depth of the defect below metal surface more exactly.

Thus, we have demonstrated the possibility of calculations of
anisotropic conductance oscillation caused by electron scattering
by the defect in noble metals. The developed algorithm of
calculations can be used for any parametrically given FS. We have
shown that the analysis of interference patterns makes it possible
to find the position of the defect below metal surface.

This work was partly supported by Fundamental Research State  Fund
of Ukraine (project $\Phi$ 25.2/122).


\begin{thebibliography}{99}
\bibitem{Avotina2} Ye. S. Avotina, Yu. A. Kolesnichenko, A.F. Otte, and J.M.
Ruitenbeek, Phys. Rev. B, \textbf{74}, 085411 (2006).

\bibitem{Cromme} M. F. Crommie, C. P. Lutz, and D. M. Eigler, Science
\textbf{262}, 218 (1993).

\bibitem{Koles} O. Yu. Kolesnychenko, R. de Kort, M. I. Katsnelson, A. I.
Lichtenstein and H. van Kempen, Nature \textbf{415}, 507 (2002).

\bibitem{Knorr} N. Knorr, H. Brune, M. Epple, A. Hirstein. M.A. Schneider, and K. Kern, Phys. Rev., \textbf{65}, 115420 (2002).

\bibitem{Burge} L. B\"{u}rgi, N. Knorr, H. Brune, M.A. Schneider, K. Kern
Appl. Phys. A \textbf{75}, 141 (2002).

\bibitem{babble} M. Schmid, W. Hebenstreit, P. Varga and S. Crampin, Phys.
Rev. Lett., \textbf{76}, 2298 (1996).

\bibitem{Quaas} N. Quaas, PhD thesis, G\"{o}ttingen University (2003).

\bibitem{Wend} N. Quaas, M. Wenderoth, A. Weismann, R.G. Ulbrich and K. Sch%
\"{o}nhammer, Phys. Rev. B \textbf{69}, 201103(R) (2004).

\bibitem{Ber} P.T. Sprunger, L. Petersen, E.W. Plummer, E. Lagsgaard, F.
Besenbacher, Science \textbf{275}, 1764 (1997).

\bibitem{Fourier} Ph. Hofmann, B.G. Briner, M. Doering, H.-P. Rust, E.W.
Plum-mer, and A.M. Bradshaw, Phys. Rev. Lett. \textbf{79}, 265 (1997).

\bibitem{Petersen} L. Petersen, P. Laitenberger, E. L\ae gsgaard, and F.
Besenbacher, Phys. Rev. B \textbf{58}, 7361 (1998).

\bibitem{Avotina1} \bigskip Ye. S. Avotina, Yu. A. Kolesnichenko, A.N.
Omelyanchouk, A.F. Otte, and J.M. Ruitenbeek, Phys. Rev. B \textbf{71},
115430 (2005).

\bibitem{Avotina3} Ye. S. Avotina, Yu. A. Kolesnichenko, A.F. Otte, and J.M.
Ruitenbeek, Phys. Rev. B \textbf{75}, 125411 (2007).

\bibitem{KMO} I. O. Kulik, Yu. N. Mitsai, and A. N. Omelyanchouk, Zh. Exp.
Teor. Fiz., \textbf{63}, 1051 (1974).

\bibitem{Kosevich} A.M. Kosevich, Fiz. Nizk. Temp., \textbf{11}, 1106 (1985)
[Sov. J. Low Temp. Phys., \textbf{11}, 611 (1985)].

\bibitem{LAK} I.M. Lifshits, M.Ya. Azbel', and M.I. Kaganov, \ "Electron
theory of metals", New York, Colsultants Bureau (1973).

\bibitem{Korn} G. Korn, T.Korn, "Mathematical Handbook", McGraw-Hill Company
(1968).

\bibitem{FS} Arthur P. Cracknell, The Fermi surfaces of metals: a
description of the Fermi surfaces of the metallic elements, Taylor and
Francis, London (1971).
\end{thebibliography}
\end{document}